\documentclass[11pt]{article}\pdfoutput=1
\usepackage{cite}
\usepackage{amsmath,amsfonts,amssymb}
\usepackage[small,bf,hang]{caption}
\usepackage{latexsym,epsfig}
\usepackage{arydshln}

\usepackage{chngcntr}

\def\hybrid{
        \topmargin -20pt
        \oddsidemargin 0pt
        \headheight 0pt \headsep 0pt
        \textwidth 6.25in 
        \textheight 9.5in 
        \marginparwidth .875in
        \parskip 5pt plus 1pt \jot = 1.5ex}

\hybrid

 \csname
@addtoreset\endcsname{equation}{section}

\def\moth{\mathsurround=0pt}
\newdimen\zo \zo=0pt

\def\tick{\leaders\hrule height 0.5ex depth 0pt \hskip 0.5pt}
\def\upboxfill{$\moth \setbox\zo\hbox{\tick}%
  \hskip 3pt\hbox to 0pt{$\tick$\hss}\hrulefill \hbox to 7.5pt{$\tick$\hss}$}

\def\dtick{\leaders\hrule height .34pt depth 0.5ex \hskip 0.5pt}
\def\downboxfill{$\moth \setbox\zo\hbox{\dtick}%
  \hskip 2pt\hbox to 0pt{$\dtick$\hss}\hrulefill \hbox to 2pt{$\dtick$\hss}$}

\def\bec{\begin{center}}
\def\ec{\end{center}}

\def\be{\begin{equation}}
\def\ee{\end{equation}}
\def\bea{\begin{eqnarray}}
\def\eea{\end{eqnarray}}
\def\ba{\begin{array}}
\def\ea{\end{array}}

\begin{document}

\begin{titlepage}
\rightline{}
\rightline{Submitted March 23, 2018}
\begin{center}
\vskip 2cm
{\Large \bf{On Background Independence in String Theory}
}\\
\vskip 2.2cm

{\large\bf {Olaf Hohm}}
\vskip 1.6cm
{\it 
Simons Center for Geometry and Physics, Stony Brook University,\\
Stony Brook, NY 11794-3636, USA}\\
ohohm@scgp.stonybrook.edu
\vskip .2cm

\end{center}

\bigskip\bigskip
\begin{center} 
\textbf{Abstract}

\end{center} 
\begin{quote}

I discuss various aspects of background independence in the context of string theory, 
for which so far we have no manifestly background independent formulation. 
After reviewing the role of background independence in classical Einstein gravity, 
I discuss recent results implying that there is a conflict in string theory between manifest 
background independence and manifest duality invariance when higher-derivative 
corrections are included. The resolution of this conflict requires 
the introduction of new gauge degrees of freedom together with an enlarged gauge symmetry. 
This suggests more generally that a manifestly background 
independent and duality invariant formulation of string theory requires significantly enhanced 
gauge symmetries.

\end{quote}

\bigskip\bigskip\bigskip\bigskip

Essay written for the Gravity Research Foundation 2018 Awards for Essays on Gravitation
 
\vfill

\end{titlepage}



\counterwithout{equation}{section}

\setcounter{equation}{0}

Einstein's theory of general relativity 
is \textit{manifestly background independent}: no background geometry has to be 
specified in order to formulate the theory. While some quantum gravity researchers have assigned 
an almost mystical significance to this fact, others dismiss its relevance  altogether.  
It is my purpose in this essay, first, to give a sober discussion of the role of background independence 
in general relativity and string theory more generally and, second, to explain recent results 
exhibiting a conflict between manifest background independence and manifest duality invariance 
that arises in `stringy gravity' with higher-derivative corrections.  
This result has potentially far-reaching implications, suggesting that any manifestly background independent 
formulation of string theory must exhibit a much larger gauge symmetry than is apparent 
in its low-energy, general relativistic description.

I begin by recalling Einstein gravity from the physical 
viewpoint advertised by Feynman, Weinberg and others \cite{Feynman:1996kb}. One starts from a  massless 
spin-2 field in Minkowski space, which is a symmetric 
rank-two tensor $h_{\mu\nu}$ under Lorentz transformations, 
subject  to the gauge redundancy 
   \be\label{gaugetrans}
    h_{\mu\nu} \;\rightarrow \; h_{\mu\nu} + \partial_{\mu}\xi_{\nu}+\partial_{\nu}\xi_{\mu}\;. 
   \ee  
Gauge invariance uniquely determines the two-derivative 
theory, which at the quadratic level is given by the Fierz-Pauli Lagrangian 
  \begin{equation}\label{FPaction}
    {\cal L}[h] \ = \ \frac{1}{2} \partial^{\mu}h^{\nu\rho} \partial_{\mu} h_{\nu\rho}
    -\partial_{\mu}h^{\mu\nu} \partial^{\rho} h_{\rho\nu} +\partial_{\mu}h^{\mu\nu}\partial_{\nu}h
    -\frac{1}{2} \partial^{\mu}h \,\partial_{\mu}h\;, 
   \end{equation}
where $h=\eta^{\mu\nu}h_{\mu\nu}$, with Minkowski metric $\eta_{\mu\nu}$.    
One may now include interactions, governed by a 
coupling constant $\kappa$ proportional to the square root of Newton's constant, by adding 
cubic terms in $h$ to (\ref{FPaction}) and linear terms in $h$ to (\ref{gaugetrans}), both of order $\kappa$. 
This iterative procedure, which can be extended to higher order in $\kappa$, never terminates, 
leading to the famously involved non-polynomiality of perturbative gravity. 
This perturbative formulation is \textit{background dependent}, because the 
background Minkowski metric $\eta_{\mu\nu}$ is needed in order to write (\ref{FPaction}).

While it is possible in principle to extend the iterative procedure of constructing the interactions 
to arbitrary orders in $\kappa$, it is certainly not practical or conceptually satisfactory. 
We have, of course, a closed-form description, which was the original formulation due to Einstein and Hilbert. 
It is recovered by introducing the \textit{background independent} field variable obtained by adding 
the spin-2 field to the Minkowski metric:  
   \begin{equation}\label{metricexpand}
    g_{\mu\nu}(x) \ \equiv \ \eta_{\mu\nu} + \kappa h_{\mu\nu}(x)\,, 
   \end{equation}
 for which the gauge transformations (\ref{gaugetrans}) can be extended to all orders
 in terms of general coordinate transformations (diffeomorphisms).  
One can then construct a Ricci scalar $R$ for 
the full metric $g_{\mu\nu}$ and write the Einstein-Hilbert action 
 \begin{equation}\label{EH} 
  S \ = \ \frac{1}{2\kappa^2} \int {\rm d}^4x\,\sqrt{-g}\,R\;, 
 \end{equation}
which yields (\ref{FPaction}) at the quadratic level.

 Let me summarize and emphasize several key features  of the above narrative 
 relating the spin-2 field theory to Einstein's geometric theory of gravity: 
 \begin{itemize}
 \item The action  (\ref{EH}) is manifestly background independent, being formulated entirely in 
 terms of the fundamental field $g_{\mu\nu}$, without any background metric.  
 Thus, the Einstein-Hilbert theory really solves two problems: 
 \textit{i)} it gives an all-order description of a massless spin-2 theory in Minkowski space, 
 and \textit{ii)} it gives a theory 
 that is valid for arbitrary, generally curved  background geometries (subject to  
 the field equations) in that we may 
 now expand (\ref{EH}) about any such background metric $\bar{g}_{\mu\nu}(x)$. 
 \item For physical applications it is typically necessary to fix a background metric and to work 
 with a perturbative formulation. Examples include: \textit{i)} the theory of gravitational 
 waves that starts with (\ref{FPaction}); \textit{ii)} cosmological perturbation theory
  \cite{Mukhanov:1990me}, 
 which is the theoretical tool of modern precision cosmology. 
 The advantage of the universal background 
 independent formulation (\ref{EH}) is simply that it can serve as starting point for arbitrary backgrounds.  
 \item
 The spin-2 theory is formulated 
 with gauge redundancy (\ref{gaugetrans}) because we are demanding  \textit{manifest Lorentz invariance}. 
 Starting from the covariant background independent formulation, 
 the local gauge invariance and global Lorentz symmetry 
 are a consequence of general coordinate invariance. 
 A formulation without gauge redundancies exists in light-cone gauge \cite{Goroff:1983hc}, 
 but then spacetime covariance and  
 locality are no longer manifest,  and this formulation does not seem useful for most applications.

 \end{itemize}

\bigskip

After this reminder, we now turn to string theory. 
We do not know of a background independent formulation of string theory, but we have closed string field theory, 
which is formulated with respect to a `background' conformal field theory that encodes in particular 
the background metric \cite{Zwiebach:1992ie}. The resulting target space actions extend (\ref{FPaction}) by 
an infinite number of component fields. 
Importantly, there is a vast space of consistent backgrounds, sometimes referred to as 
the \textit{string landscape}, which some have used to argue that string theory is not predictive, 
since we do not know which part of the landscape we inhabit. This 
criticism is misguided, however, because general relativity itself features a landscape, yet is 
perfectly predictive. Here the landscape consists of 
all metrics $\bar{g}_{\mu\nu}$  satisfying the vacuum Einstein equations. 
If anything, the landscape of general relativity is incomparably larger, carrying a 
continuous infinity of backgrounds, while in string theory (subject to further consistency 
conditions such as `flux quantization') this number may be even finite. 
Arguably, the real challenge of string theory is then to find a background independent formulation 
and a `covariance' principle that would allow one to apply and test it in a manner similar to 
general relativity.

In the following I will mimic the above logic of deriving general relativity 
from the massless spin-2 theory for the universal massless fields of string theory: 
a rank-two tensor $h_{\mu\bar{\nu}}$, combining the symmetric graviton with 
an antisymmetric (Kalb-Ramond) field, and a scalar (dilaton) $\Phi$. 
Their quadratic Lagrangian on flat space reads \cite{Hull:2009mi,Siegel:1993th}
   \begin{equation}\label{DFTlin} 
   {\cal L} \ = \ \frac{1}{2}\Big(D^{\mu}h^{\nu\bar{\rho} } D_{\mu}h_{\nu\bar{\rho}} \ - \  
   D^{\mu}h^{\nu\bar{\rho}}
  D_{\nu}h_{\mu\bar{\rho}} \ + \ \bar{D}^{\bar{\mu}}h^{\nu\bar{\rho}} \bar{D}_{\bar{\rho}}h_{\nu\bar{\mu} }
  \ - \ 2\bar{D}_{\bar{\nu}}h^{\mu\bar{\nu}} D_{\mu}\Phi \ + \ D^{\mu}\Phi D_{\mu}\Phi\Big)\;, 
  \end{equation} 
with two types of indices, $\mu,\nu = 1,\ldots, d$, $\bar{\mu},\bar{\nu} = 1,\ldots, d$, 
and differential operators w.r.t.~doubled coordinates 
$X=(x,\tilde{x})$:  
 \begin{equation}
   D \ \equiv \ \partial  \ - \  E\,\tilde{\partial}\,, \qquad
   \bar{D} \ \equiv \ \partial \ + \ E^t\,\tilde{\partial}\;. 
  \end{equation}
The constant matrix $E$ encodes the sum of background metric and 
Kalb-Ramond field. In (\ref{DFTlin}) all unbarred and barred indices are consistently 
contracted, which implies a doubled `Lorentz' invariance under  
  \begin{equation}\label{doubledLorentz}
   h_{\mu\bar{\nu}}(X) \; \rightarrow \; h^{\prime}_{\mu\bar{\nu}}(X') \ = \  M_{\mu}{}^{\rho}\, 
   \bar{M}_{\bar{\nu}}{}^{\bar{\sigma}} \, h_{\rho\bar{\sigma}}(X)
    \;\;,\qquad 
   M \ \in \ {SO}(d)_L\;, \;\; \bar{M} \ \in \ {SO}(d)_R\;, 
  \end{equation} 
here written for euclidean signature. The action is invariant under 
diffeomorphisms generalizing (\ref{gaugetrans}),  with parameters $\xi_{\mu},\, \bar{\xi}_{\bar{\mu}}$: 
  \begin{equation}\label{gengauge}
   \delta h_{\mu\bar{\nu}} \ =  \ D_{\mu} \bar{\xi}_{\bar{\nu}} \ + \ \bar{D}_{\bar{\nu}} \xi_{\mu} \,, \qquad
   \delta \Phi \ = \ D_{\mu}\xi^{\mu} + \bar{D}_{\bar{\mu}}\bar{\xi}^{\bar{\mu}}\;. 
  \end{equation}  
In string theory, the doubling of coordinates is due 
to winding modes on toroidal backgrounds (formally, (\ref{DFTlin}) remains valid 
for non-compact flat backgrounds), subject to the so-called level-matching constraint 
$D^{\mu}D_{\mu}=\bar{D}^{\bar{\mu}}\bar{D}_{\bar{\mu}}$. 
In addition, the action is invariant  
under the larger group $O(d,d)$, provided the background transforms as 
 \be
  E \;\rightarrow \; (aE+b)(cE+d)^{-1}\;, \qquad 
  \begin{pmatrix}   a & b\\
  c & d \end{pmatrix}  \ \in \ O(d,d)\;, 
 \ee
which includes the T-duality inversion of radii, $R\rightarrow {\alpha'}/{R}$.  

In the same way that we asked above for a background independent theory 
with a covariance principle that implies gauge and Lorentz invariance when expanding 
about Minkowski space, we now ask for a manifestly background independent theory that guarantees 
the doubled `Lorentz' symmetry (and the full $O(d,d)$ duality) upon expansion about 
flat space. For the two-derivative theory there is a compelling answer, double field theory, 
which can be obtained, as in (\ref{metricexpand}), by introducing  a background independent field, 
the \textit{generalized metric} 
   \begin{equation}\label{genmetrc}
   {\cal H}_{MN} \ = \   \begin{pmatrix}    g^{\mu\nu} & -g^{\mu\rho}b_{\rho\nu}\\[0.5ex]
  b_{\mu\rho}g^{\rho\nu} & g_{\mu\nu}-b_{\mu\rho}g^{\rho\sigma}b_{\sigma\nu}
  \end{pmatrix} \;, 
  \end{equation}
where  $M, N=1,\ldots, 2d$ are fundamental $O(d,d)$ indices. 
The generalized metric is $O(d,d)$ valued, and its fluctuations can be parametrized 
in terms of a tensor  $h_{\mu\bar{\nu}}$. 
There is a generalized notion of diffeomorphisms, extending (\ref{gengauge}) to all orders, 
and a generalized notion of geometry that allows one to define a generalized curvature scalar 
${\cal R}$ and thus an Einstein-Hilbert-type action \cite{Hohm:2010jy,Hohm:2010pp}: 
  \begin{equation}\label{fullDFT}
   S \ = \ \int {\rm d}X\, e^{\Phi}\,{\cal R}({\cal H}, \Phi)\;. 
  \end{equation}

We have succeeded in finding a manifestly background independent and duality invariant 
formulation at the two-derivative level, but in string theory  there are also higher derivative 
corrections to (\ref{DFTlin}), governed by the dimensionful (inverse) string tension $\alpha'$. 
Is there a background independent extension of (\ref{DFTlin}) 
encoding not only higher orders in fields (in $\kappa$), but also in $\alpha'$? The answer is affirmative 
and at the same time more involved and more intriguing than anyone anticipated: 
the inclusion of $\alpha'$ corrections requires an $\alpha'$-deformation 
of the gauge structure. 

To first order in $\alpha'$, and cubic order in fields, string field theory yields 
the following deformation of the gauge transformations \cite{Hohm:2014xsa}:  
  \begin{equation}\label{deformedgauge}
    \Delta_{\xi} h_{\mu\bar{\nu}} \ = \  \alpha'
    \Big( D_{\mu}
    D^{\rho}\xi^{\sigma}
    \,\Gamma_{\bar{\nu}\rho\sigma} 
    \ - \ \bar{D}_{\bar{\nu} }\bar{D}^{\bar{\rho}}\bar{\xi}^{\bar{\sigma}}\,\Gamma_{\mu\bar{\rho}\bar{\sigma}}\Big)
    \ + \ {\cal O}(\alpha^{\prime 2}) \;, 
  \end{equation}  
where
    \begin{equation}
  \Gamma_{\mu\bar{\nu}\bar{\rho}} \ \equiv  \ \bar{D}_{\bar{\nu}}h_{\mu\bar{\rho}} 
  - \bar{D}_{\bar{\rho}}h_{\mu\bar{\nu}}\;, 
  \qquad 
  \Gamma_{\bar{\mu}\nu\rho} \ \equiv  \ D_{\nu}h_{\rho\bar{\mu}} -  D_{\rho}h_{\nu\bar{\mu}}\;. 
 \end{equation}  
This deformation is non-trivial: it cannot be removed by \textit{duality covariant} 
redefinitions. (It can be removed by \textit{duality violating} redefinitions, but this, 
of course, defeats the purpose).

Surprisingly, it can be proved that the transformation (\ref{deformedgauge}) 
\textit{cannot} be obtained from background independent $\alpha'$-deformed gauge 
transformations of the generalized metric \cite{Hohm:2016lge,Hohm:2016yvc}. 
We may describe this result as follows: \\[1ex]
\textit{There is a conflict between manifest background independence and manifest duality invariance
once higher-derivative $\alpha'$ corrections are included.} \\[-1.5ex]

This conflict can be resolved by using   
a frame (vielbein) formulation. Conventionally, a formulation with 
frame $E_{A}{}^{M}$  defines a generalized metric 
as 
 \begin{equation} 
  {\cal H}_{MN} \ = \ E_M{}^{A} E_{N}{}^{B}\eta_{AB}\;, 
 \end{equation} 
  with `tangent space' metric $\eta_{AB}$, 
and is thus, in absence of fermions, equivalent to a generalized metric formulation. The frame $E_{A}{}^{M}$ 
encodes more component fields but is subject to the local frame transformations 
$\delta_{\Lambda}E_{A}{}^{M}=\Lambda_A{}^{B} E_{B}{}^{M}$, $\Lambda\in SO(d)_L\times SO(d)_R$, 
which render the unphysical degrees of freedom pure gauge. 
The crucial observation is now that (\ref{deformedgauge}) 
\textit{can} be obtained from $\alpha'$-deformed frame transformations,  
which are background independent but for which the generalized metric is \textit{not}  
an invariant object. 
The unphysical degrees of freedom encoded in $E_{A}{}^{M}$ may still be gauged away,  
but \textit{only upon fixing a background} \cite{Hohm:2016lge,Hohm:2016yvc,Marques:2015vua}.

Summarizing, a manifestly background independent and duality invariant formulation of string theory 
including $\alpha'$ corrections 
requires an enhanced gauge symmetry (in the form of frame transformations). 
The consistency of this $\alpha'$-deformed geometry has so far only 
been established to first order in $\alpha'$, and it is plausible that to higher order in $\alpha'$ 
we may have to enhance the gauge symmetry further, 
and possibly include even `higher gauge structures' \cite{Hohm:2017pnh}.
 This suggests  
that the gauge symmetry of `stringy gravity' may have to be much larger 
than is apparent from the usual `Einsteinian' formulation.  
While it has already been argued from different angles that string theory ultimately may exhibit
a much larger gauge symmetry \cite{Gross:1988ue,Witten:1988sy}, 
I believe that the viewpoint advanced here is novel and 
thus provides a unique opportunity to bring us closer to the elusive principles of string theory 
and quantum gravity.

\medskip

 \section*{Acknowledgments} 

I would like to thank Hermann Nicolai, Ashoke Sen and Barton Zwiebach for discussions. 
This research was supported by a Heisenberg Fellowship 
of the German Science Foundation (DFG).


\begin{thebibliography}{99}

\bibitem{Feynman:1996kb} 
  R.~P.~Feynman, F.~B.~Morinigo, W.~G.~Wagner and B.~Hatfield,
  ``Feynman lectures on gravitation,''
  Reading, USA: Addison-Wesley (1995). 
  
\bibitem{Mukhanov:1990me} 
  V.~F.~Mukhanov, H.~A.~Feldman and R.~H.~Brandenberger,
  ``Theory of cosmological perturbations. Part 1. Classical perturbations. Part 2. Quantum theory of perturbations. Part 3. Extensions,''
  Phys.\ Rept.\  {\bf 215}, 203 (1992).
  doi:10.1016/0370-1573(92)90044-Z
  
\bibitem{Goroff:1983hc} 
  M.~Goroff and J.~H.~Schwarz,
  ``$D$-dimensional Gravity in the Light Cone Gauge,''
  Phys.\ Lett.\  {\bf 127B}, 61 (1983).
  doi:10.1016/0370-2693(83)91630-1

\bibitem{Zwiebach:1992ie} 
  B.~Zwiebach,
  ``Closed string field theory: Quantum action and the B-V master equation,''
  Nucl.\ Phys.\ B {\bf 390}, 33 (1993)
  doi:10.1016/0550-3213(93)90388-6
  [hep-th/9206084].
  
\bibitem{Hull:2009mi} 
  C.~Hull and B.~Zwiebach,
  ``Double Field Theory,''
  JHEP {\bf 0909}, 099 (2009)
  doi:10.1088/1126-6708/2009/09/099
  [arXiv:0904.4664 [hep-th]].  

\bibitem{Siegel:1993th} 
  W.~Siegel,
  ``Superspace duality in low-energy superstrings,''
  Phys.\ Rev.\ D {\bf 48}, 2826 (1993)
  doi:10.1103/PhysRevD.48.2826
  [hep-th/9305073].
  
\bibitem{Hohm:2010jy} 
  O.~Hohm, C.~Hull and B.~Zwiebach,
  ``Background independent action for double field theory,''
  JHEP {\bf 1007}, 016 (2010)
  doi:10.1007/JHEP07(2010)016
  [arXiv:1003.5027 [hep-th]].  
  
\bibitem{Hohm:2010pp} 
  O.~Hohm, C.~Hull and B.~Zwiebach,
  ``Generalized metric formulation of double field theory,''
  JHEP {\bf 1008}, 008 (2010)
  doi:10.1007/JHEP08(2010)008
  [arXiv:1006.4823 [hep-th]].  

\bibitem{Hohm:2014xsa} 
  O.~Hohm and B.~Zwiebach,
  ``Double field theory at order $\alpha'$,''
  JHEP {\bf 1411}, 075 (2014)
  doi:10.1007/JHEP11(2014)075
  [arXiv:1407.3803 [hep-th]].
  
\bibitem{Hohm:2016lge} 
  O.~Hohm,
  ``Background Independence and Duality Invariance in String Theory,''
  Phys.\ Rev.\ Lett.\  {\bf 118}, no. 13, 131601 (2017)
  doi:10.1103/PhysRevLett.118.131601
  [arXiv:1612.03966 [hep-th]]. 
  
\bibitem{Hohm:2016yvc} 
  O.~Hohm,
  ``Background Independent Double Field Theory at Order $\alpha'$: Metric vs. Frame-like Geometry,''
  Phys.\ Rev.\ D {\bf 95}, no. 6, 066018 (2017)
  doi:10.1103/PhysRevD.95.066018
  [arXiv:1612.06453 [hep-th]]. 
  
\bibitem{Marques:2015vua} 
  D.~Marques and C.~A.~Nunez,
  ``T-duality and $\alpha'$-corrections,''
  JHEP {\bf 1510}, 084 (2015)
  doi:10.1007/JHEP10(2015)084
  [arXiv:1507.00652 [hep-th]]. 
  
\bibitem{Hohm:2017pnh} 
  O.~Hohm and B.~Zwiebach,
  ``$L_{\infty}$ Algebras and Field Theory,''
  Fortsch.\ Phys.\  {\bf 65}, no. 3-4, 1700014 (2017)
  doi:10.1002/prop.201700014
  [arXiv:1701.08824 [hep-th]].  

\bibitem{Gross:1988ue} 
  D.~J.~Gross,
  ``High-Energy Symmetries of String Theory,''
  Phys.\ Rev.\ Lett.\  {\bf 60}, 1229 (1988).
  doi:10.1103/PhysRevLett.60.1229

\bibitem{Witten:1988sy} 
  E.~Witten,
  ``The Search For Higher Symmetry In String Theory,''
  Phil.\ Trans.\ Roy.\ Soc.\ Lond.\ A {\bf 329}, 349 (1989).

 \end{thebibliography}
\end{document}